\documentclass[aps,prb,preprint,groupedaddress]{revtex4-1}

\usepackage{graphicx}
\usepackage{natbib}

\begin{document}


\title{Coupling of Low Energy Electrons in Optimally Doped Bi$_{2}$Sr$_{2}$CaCu$_{2}$O$_{8+\delta}$ to an Optical Phonon Mode}


\author{J.D. Rameau}
\author{H.-B. Yang}
\author{G.D. Gu}
\author{P.D. Johnson}
\affiliation{Condensed Matter Physics and Materials Science Department
Brookhaven National Laboratory
Upton, NY 11973}


\date{\today}

\begin{abstract}
Laser based photoemission with photons of energy 6 eV is used to examine the fine details of the very low energy electron dispersion and associated dynamics in the nodal region of optimally doped Bi2212. A "kink" in the dispersion in the immediate vicinity of the Fermi energy is associated with scattering from an optical phonon previously identified in Raman studies. The identification of this phonon as the appropriate mode is confirmed by comparing the scattering rates observed experimentally with the results of calculated scattering rates based on the properties of the phonon mode.
\end{abstract}


\maketitle

\section{Introduction}
In the last decade, with increased energy and momentum resolution, angle resolved photoemission spectroscopy (ARPES) has emerged as a powerful probe of the mass renormalizations, or "kinks", in the electron spectral function associated with the coupling of electrons to collective excitations in condensed matter systems.\cite{ref1}  In particular in many systems it has been possible to clearly identify the coupling to phonons, spin waves and charge density waves.\cite{ref1}  In the high $T_{c}$ cuprates on the other hand the identification of the modes responsible for the observed mass renormalizations has often proved controversial, particularly when different modes occur at the same energy.  Experimentally a mass renormalization and associated kink in the dispersion at 70 meV \cite{ref2} has been attributed to either spin excitations \cite{ref3} or to phonons \cite{ref4}.   However a recent theoretical study \cite{ref5} has suggested that coupling to phonons in the 70 meV energy range and above is, by itself, too small to produce the observed mass renormalization.  A mass renormalization observed at 350 meV has also attracted considerable attention but again that exact region remains controversial \cite{ref6}. In the present communication, using laser based photoemission \cite{ref7a}\cite{ref7b} we present evidence for an additional kink in the nodal region of optimally doped Bi$_{2}$Sr$_{2}$CaCu$_{2}$O$_{8+\delta}$ (Bi2212,$T_{c}$=91 K).  We demonstrate that this new kink at low energies exhibits a unique dispersion with respect to the superconducting gap in the nodal region of the Brillouin zone. That is, it always appears ~8 meV below the superconducting gap regardless of the gap's magnitude. Because of this behavior we can unequivocally associate an observed mass renormalization in a high $T_{c}$ cuprate with a single phonon mode.

\section{Experimental Methodology}

The ARPES study was carried out with a laser based 6 eV photon source produced as the fourth harmonic of a mode-locked Ti:Sapphire oscillator. The oscillator emitted transform limited pulses of ~2 ps duration at a repetition rate of 105 MHz. The spectral width of the fourth harmonic beam was confirmed to be less than or equal to 2.5 meV FWHM with a deep ultraviolet imaging spectrometer at the limit of its resolving power. The 6 eV photon beam was focused onto the sample with a 100 mm focal length VUV grade CaF$_{2}$ lens. The sample, a high quality, single crystal of optimally doped Bi2212, was cleaved in situ at a temperature of 10 K. All data were acquired at the same 10 K within five hours of exposing the fresh crystal surface. The chemical potential was referenced to a pure gold sample in electrical contact with the Bi2212 crystal. Photoemission spectra were recorded with a Scienta SES-2002 spectrometer with an overall energy resolution in the present experiment of 13 meV and a momentum resolution of ~.003 {\AA}$^{-1}$ at the photon energies used.

\begin{figure}
  \includegraphics[scale = .5, bb = -146 116 626 705]{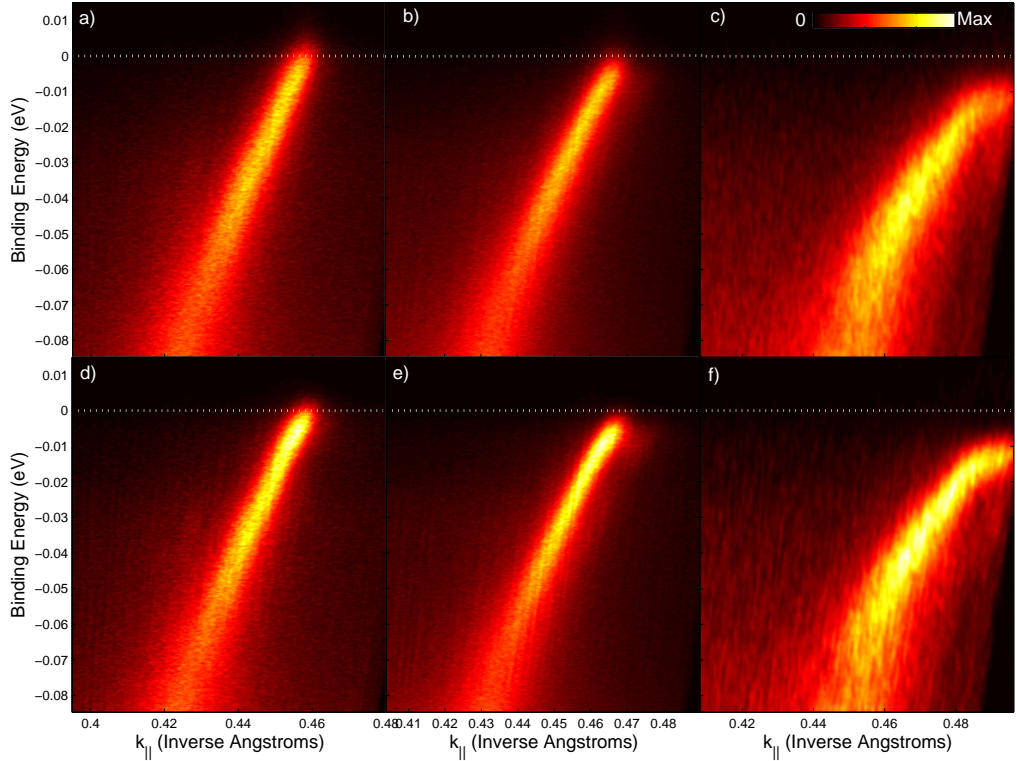}\\
  \caption{(Color online) False color intensity maps of the raw (panels a-c) and deconvolved (panels d-f) ARPES spectra.  The color scaling is depicted at the top of panel c. The intensity scales of the raw spectra are normalized to those of their deconvolved counterparts.}\label{fig1}
\end{figure}

\begin{figure}
  \includegraphics[scale = .8, bb = 133 210 468 666]{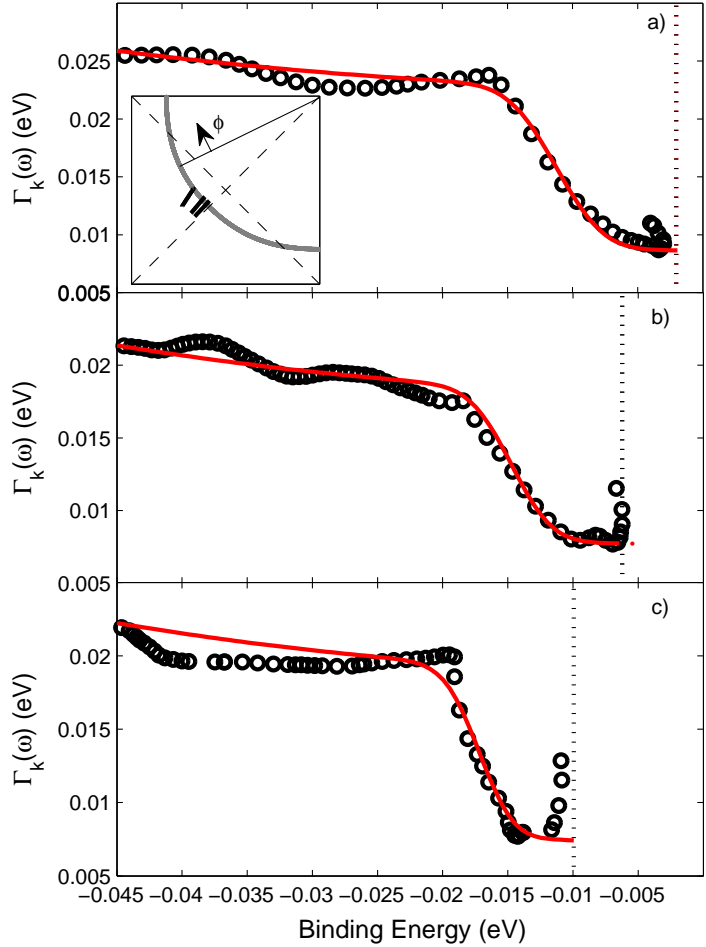}\\
  \caption{(Color online) Panels a-c show scattering rates $\Gamma_{\vec{k}}(\omega)$ extracted directly from EDC's of the spectra shown in Figure \ref{fig1}d-f, respectively. The dotted lines denote the gap energy relative to the midpoint step energy. Red lines are fits to Eq. \ref{lam} with $\alpha^{2}$=1.5, 1.1 and 1.2 including resolution broadening, impurity scattering and the $\omega_{kink}=\Delta_{\vec{k}}+\Omega_{0}$ dependence of the kink energy. The inset of panel a diagrams the locations of cuts in the Brillouin zone. The gap values shown in panels a and c have been corrected for the finite energy resolution and kinematic effects, respectively.}\label{fig2}
\end{figure}

ARPES spectra acquired at three locations along the Fermi surface are shown in Figure \ref{fig1}(a-c). The location of these cuts in the Brillouin zone is shown in the inset of Figure \ref{fig2}a. It has been shown previously that at a photon energy of 6 eV the antibonding band reflecting the bilayer splitting is entirely absent from the spectra \cite{ref8}.  As discussed below the application of the Lucy-Richardson method (LRM) \cite{ref9} reduces the broadening due to the instrumental resolution, thereby sharpening the spectra considerably.  This is illustrated in figures \ref{fig1} (d-f).  Following such an analysis the effective energy resolution in the experiment was of the order of 5 meV and more importantly for these studies the effective momentum resolution was of the order of 0.001 {\AA}$^{-1}$.

\section{Experimental Analysis: The Lucy-Richardson Deconvolution}

As has been pointed out previously \cite{ref9} ARPES spectra acquired with imaging analyzers such as the Scienta 2002 used in the present experiment suffer from a non-trivial finite experimental energy and angular resolution. The finite resolution of an ARPES experiment manifests itself in the fact that the spectra acquired, e.g. Figure \ref{fig1}a-c, are an intensity map $I(\vec{k},\omega)\propto R(\vec{k},\omega|\vec{k}',\omega') \otimes A(\vec{k}',\omega')f(\omega')$ where $A$ is the underlying spectral function describing the distribution of electronic states in momentum and energy in a sample, $f(\omega')$ is the Fermi-Dirac distribution for a given temperature and $R$ is a kernel describing the resolution broadening in momentum and energy that obscures the spectral function. This convolution is responsible for generating the broadening observed in a real ARPES spectrum, $I$. Here the proportionality denotes transition matrix element effects unimportant to the analysis. $R$ is taken to be a two dimensional Gaussian distribution with full widths at half maximum parametrizing the energy and momentum resolution, respectively.

While a full description of the how the LRM as applied to ARPES will be presented elsewhere \cite{ref10}, it suffices here to recall that it is an iterative, essentially statistical "fitting" procedure that requires as it's only input the experimentally acquired spectrum $I$ and the resolution kernel $R$ which is itself routinely acquired in the course of an experiment. Thus, unlike the Maximum Entropy Method (MEM) of spectral deconvolution no \textit{a priori} assumptions as to the nature of the underlying spectral function are required to extract the spectral function. This is highly beneficial when studying phenomena with ARPES for which no underlying theory is agreed to exist. When applied to ARPES spectra the LRM produces a deconvolved image $A$ that is, statistically speaking, \textit{the most likely} physical spectral function to exist in nature that results in the measured image $I$ when convolved with the inevitable instrumental broadening $R$.

It is well known that the ill effects of finite resolution on ARPES spectra are most predominant at and around the Fermi level \cite{ref11}. Indeed, the LRM proves most valuable when called upon to correct the measured peak widths of EDC's and MDC's close to the Fermi level. This correction has the added bonus of correcting the measured Fermi velocity $v_{F}$ around the Fermi level, which tends to be radically increased in an unphysical way by resolution broadening. Unfortunately, the widths and dispersions of very low energy electrons, precisely where this problem is most acute, are also often of the most fundamental importance when comparing ARPES experiments to theory.

\begin{figure}
  \includegraphics[scale = .8, bb = 59 233 549 539]{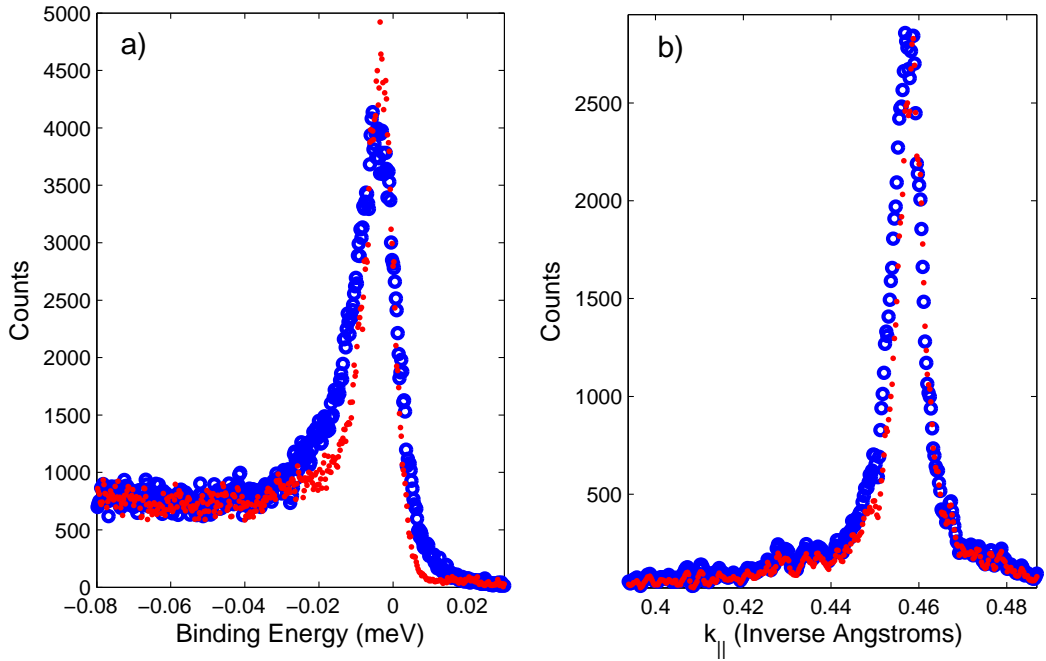}\\
  \caption{(Color online) a) EDC's at $k_{F}$ from the original data of Figure \ref{fig1}a (open circles) and deconvolved data of Figure \ref{fig1}d (dots). b) MDC's at $E_{F}$ from the original data of Figure \ref{fig1}a (circles) and deconvolved data of Figure \ref{fig1}d (dots).}\label{fig3}
\end{figure}

\begin{figure}
  \includegraphics[scale = .8, bb = 57 227 549 542]{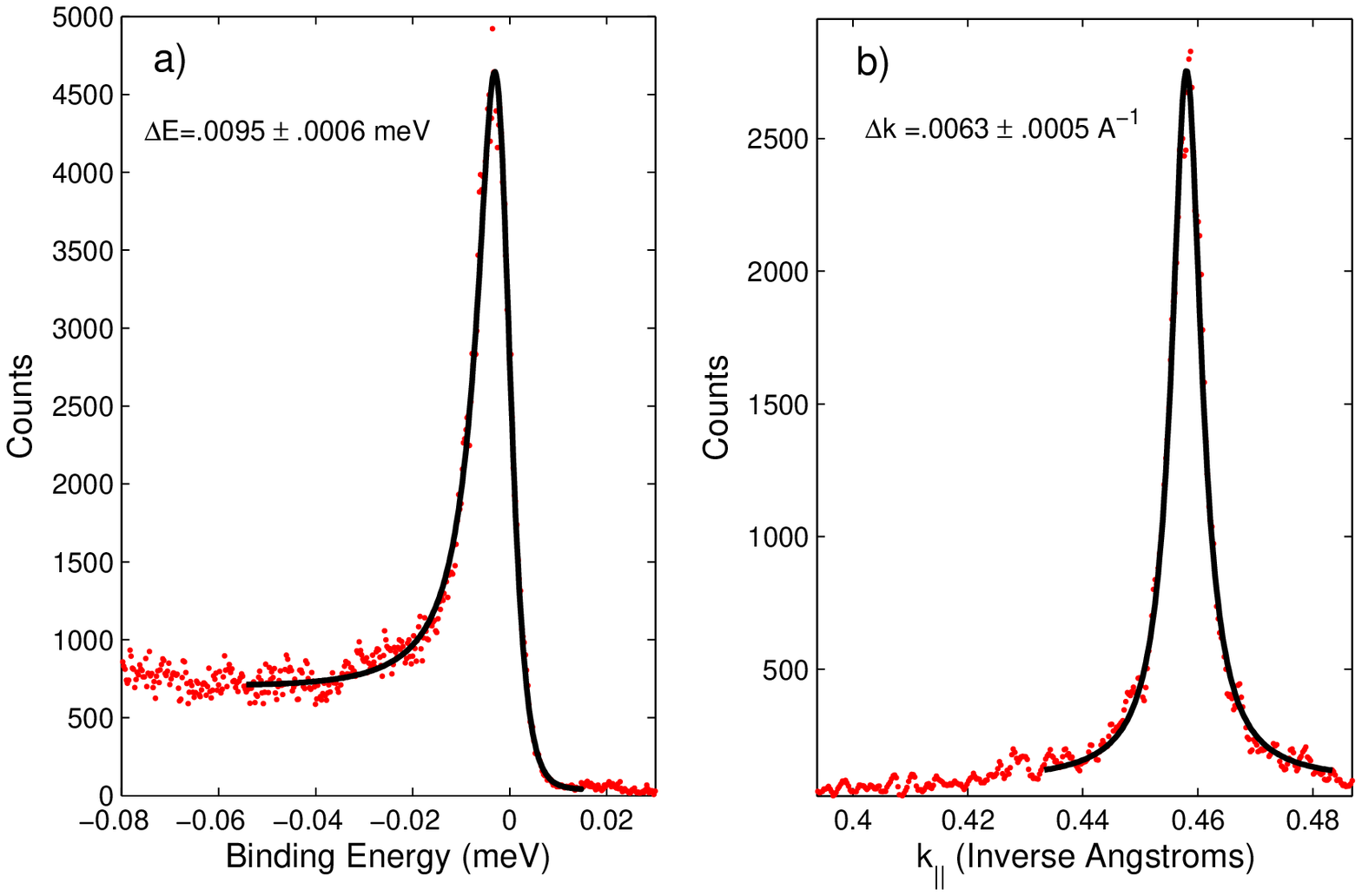}\\
  \caption{(Color online) a) Lorentzian fit (black line) to deconvolved data of Figure \ref{fig3}a including linear background and Fermi cutoff. b) Lorentzian fit with linear background to deconvolved data of Figure 3b.}\label{fig4}
\end{figure}

As an example of how the problem of finite resolution can be partially mitigated by the LRM in an ARPES experiment we plot in Figure \ref{fig3} raw and deconvolved EDC's and MDC's taken at $\vec{k}_{F}$ and $E_{F}$, respectively, from Figures \ref{fig1}a and \ref{fig1}d. The sharpening of both the EDC's and MDC's is clearly evident in the comparison. Further, the shift in the peak position of the MDC with deconvolution is indicative of the Fermi velocity also being corrected by this procedure. The widths obtained from Lorentzian fits of the deconvolved EDC and MDC are shown in Figure \ref{fig4}. The improvement of both the overall shape as well as widths of the peaks is self-evident.

\section{Analysis of Deconvolved Data}

As will be shown below, the broadening of the band at binding energies greater than the kink energy is in fact the result of an interaction between electrons in the bonding band with a well defined low energy bosonic mode. This interaction produces a renormalization of the real and imaginary parts of the electron self energy, a "kink", that itself disperses at fixed energy with respect to the bottom of the d-wave superconducting gap.

Because the observed mode sets in at such low energies and causes rapid changes in the band dispersion over a very small energy range an analysis of momentum distribution curves (MDC's) does not suffice to extract the imaginary part of the self energy.  MDC analyses were further complicated by the Bogoliubov dispersion acquired by the band with the opening of the superconducting gap.  Indeed the opening of the gap yields a double valued function in momentum up to several meV below the gap edge with a very rapidly changing, nonlinear dispersion. The spectra were thus analyzed with EDC's by fitting the spectra to Lorentzians with a linear background.  While it has been suggested that a more complicated, intrinsically asymmetric line shape is required to fit EDC's of the cuprates accurately we find that the deconvolved data at low binding energies, exemplified in Figure \ref{fig4}, are indeed very well fit at both the high and low binding energy side by simple Lorentzians.  Inverse lifetimes or scattering rates, $\Gamma_{\vec{k}}(\omega)$, extracted in this manner are shown in Figure \ref{fig2} as a function of binding energy.  At very low binding energies, in the vicinity of the gap edge, the lifetime rapidly increases and reaches a maximum at $\vec{k}\sim\vec{k}_{F}$, e.g. where the peak value of the superconducing gap $\Delta_{\vec{k}}$ is measured. Moving from left to right past $\vec{k}_{F}$ (in Figure \ref{fig1}) the lifetime decreases while the band dispersion folds back. The increase in peak width to the right of $\vec{k}_{F}$ and folding back of the dispersion reflect the redistribution of spectral weight between the occupied and unoccupied branches of the energy symmetric Bogoliubov quasiparticle dispersion that accompanies the opening of a superconducting gap. A steep increase in $\Gamma$ around 8.0 meV below the gap edges represents a second pronounced feature in the figure.  Such a sudden increase in peak width, combined with the kinks visible in the spectra in Figure \ref{fig1}d-f, are clear indications that the low energy states are interacting with a bosonic mode at a well defined energy.

\begin{figure}
  \includegraphics[scale = .8, bb = 74 213 497 562]{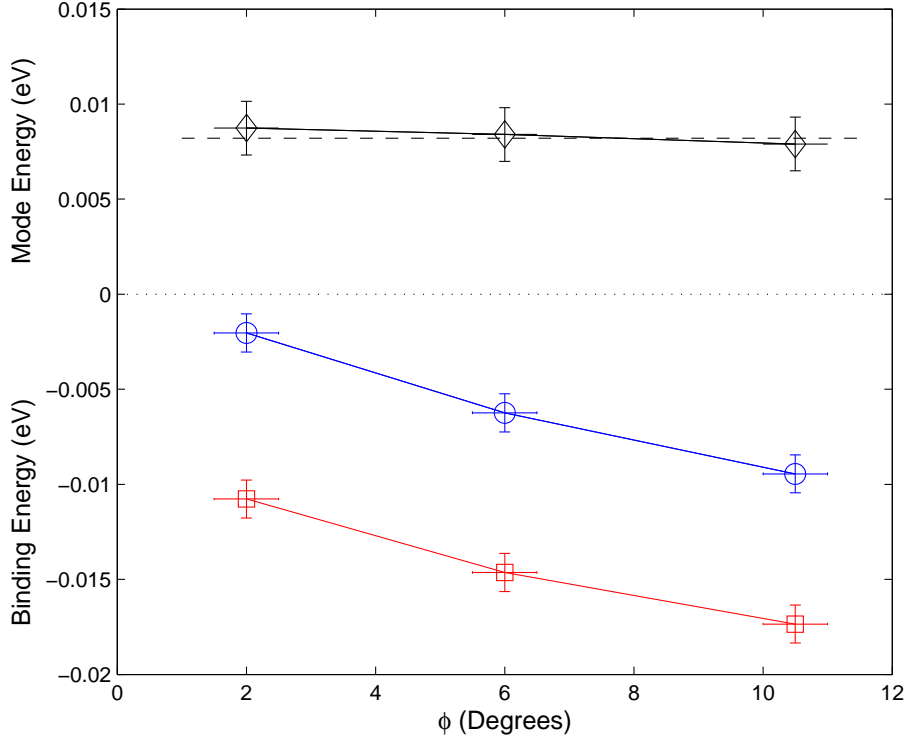}\\
  \caption{(Color online) Gap energies $\Delta_{\vec{k}}$ (circles), mode energies $\Omega_{\vec{k}}$ (squares) and the difference between the two $\delta_{\vec{k}}=\Omega_{\vec{k}}-\Delta_{\vec{k}}$ (diamonds) as a function of Fermi surface angle as defined in the inset of Figure \ref{fig2}a. }\label{fig5}
\end{figure}

We plot the mode energy, $\Omega_{\vec{k}}$, obtained from the theoretical fits described below, the superconducing gap, $\Delta_{\vec{k}}$, and the energy difference $\delta_{\vec{k}}=|\Omega_{\vec{k}}-\Delta_{\vec{k}}|$ as a function of Fermi surface angle in Figure \ref{fig5}. Remarkably it is found that over the region of Fermi surface investigated in this study $\delta_{\vec{k}}$ has a nearly constant value of ~8 meV. The gap energies, denoted by the dashed lines in Figure \ref{fig2} have in the case of panel (a) been corrected for the peak shift caused by residual resolution broadening as indicated from simulation and in the case of panel (c) by considering the Fermi surface angle of the kink location as a function of Fermi surface angle $\phi$ according to the $\Delta(\phi) = \Delta_{0} \cos(2\phi)$ parametrization of the d-wave gap with $\Delta_{0}=40$ meV. Great care must be taken in identifying the correct value of the d-wave superconducting gap for spectra acquired off of a high symmetry direction because at very low electron kinetic energies the kinematic compression effect \cite{ref12} becomes very severe.

\begin{figure}
  \includegraphics[scale = .9, bb = 140 320 471 468]{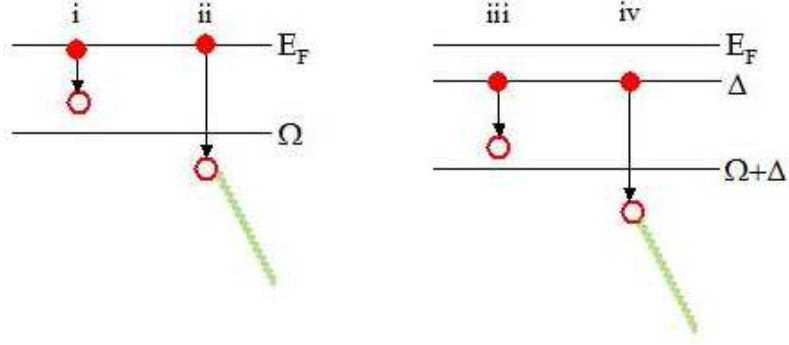}\\
  \caption{(Color online) Schematic representaton of the photohole deexcitation process. At the node a photohole (open circle) is annihilated by an electron (filled circle) from the Fermi sea (i) when the energy is less than that of the mode energy $\Omega$. When the photohole resides at a binding energy greater than $\Omega$ (ii) the decay process involves excitation of a phonon (diagonal line). Away from the node, the opening of a superconducing gap $\Delta$ in the single particle density of states shifts the absolute value of the binding energies of electrons with sufficient energy for phonon excitation in the photohole deexcitation process. The decay process in (iii) shows that electrons with energy between $\Delta$ and $\Omega+\Delta$ do not have sufficient energy to excite the mode. Photoholes at binding energies greater than $\Omega+\Delta$ (iv) are annihilated by electrons with sufficient energy to excite the mode.}\label{fig6}
\end{figure}

The fact that the step in the scattering rate is so sharp and that the value of $\delta_{\vec{k}}$ in Figure \ref{fig5} is approximately constant, even in the presence of an enlarging superconducting gap, implies zero momentum transfer or coupling to a $\vec{q}\rightarrow 0$ bosonic mode \cite{ref13}.  Physically this phenomenon derives from the fact that a photohole injected into the system at a momentum $\vec{k}$ and binding energy $\omega$ greater than $\delta_{\vec{k}}$ can, in the presence of a zone center mode, only be removed by an electron decaying in a vertical, momentum conserving transition. Because in the superconducting state all electrons below the gap energy are paired, the lowest energy electron that can transition in this manner will always originate from the vicinity of the gap edge, in the present case ~8 meV away, and so as one scans around the Fermi surface the mode energy will always appear at the same binding energy relative to the gap. This situation is illustrated schematically in Figure \ref{fig6}.

The form and binding energy of $\delta_{\vec{k}}$ observed in the present ARPES study places a strong constraint on the possible bosonic modes involved in the de-excitation.  Examination of published IR and Raman spectra \cite{ref14} \cite{ref15,ref16,ref17,ref18} indicates the presence of a very few, nearly degenerate c-axis optical phonon modes in the appropriate energy range, the most likely of which is the Raman active $A_{1g}$ mode previously observed between 58 and 65 cm$^{-1}$. In light of the theoretical work \cite{ref13} and the observed behavior of $\delta_{\vec{k}}$ we assign this optical phonon mode to the 8 meV kink observed in the APRES experiment. This observation is consistent with recent temperature dependent work at the node alone \cite{ref19} of optimally doped Bi2212. Clearly however a full theoretical understanding of the angle dependent dispersion of this kink requires more detailed work than was presented in \cite{ref13} as the kink is clearly not pinned to the maximum d-wave gap but rather the "local" gap magnitude as a function of Fermi surface angle.

Taking $\Omega_{0}$ = 8 meV and $F(\omega)$ a Gaussian of full width at half maximum ~1 meV from a fit to the Raman data we apply the standard Eliashberg equations \cite{ref20} to calculate the mass enhancement factor $\lambda$ and the scattering rate $\Gamma_{\vec{k}}(\omega)$ with $\alpha^{2}$ as a fitting parameter such that
\begin{equation}\label{lam}
    \lambda = 2\int_{0}^{\infty}\frac{\alpha^{2}F(\omega)}{\omega}d\omega
\end{equation}
\begin{equation}\label{intgam}
    \Gamma(\omega,T)=2\pi\int_{0}^{\infty}\alpha^{2}F(\omega')[2n(\omega')+f(\omega'+\omega)+f(\omega'-\omega)]d\omega'
\end{equation}
where $n(\omega')$ and $f(\omega')$ are the Bose-Einstein and Fermi-Dirac distributions, respectively. The scattering rates calculated in this fashion, shown as fits to the data in Figure \ref{fig2}, agree well with the data once shifted to account for the angular dependence of $\delta_{\vec{k}}$. The fits include additional terms reflecting a small, constant impurity scattering component (here set to 5 meV) as well as the residual energy resolution. No attempt was made in this analysis to account self consistently for the presence of the superconducting gap or the presence of the kink at 70 meV though we expect these effects to be small, excluding the phenomenological shift in binding energy of the kink. From these fits we determine $\alpha^{2}$ to be between 1.5 closest to the node and ~1.15 further away yielding $\lambda$ between 0.52 and 0.4, placing the interaction in the weak coupling regime and in good agreement with the average $\lambda$ calculated in reference \cite{ref5}.

\begin{figure}
  \includegraphics[scale = .9, bb = 89 226 482 546]{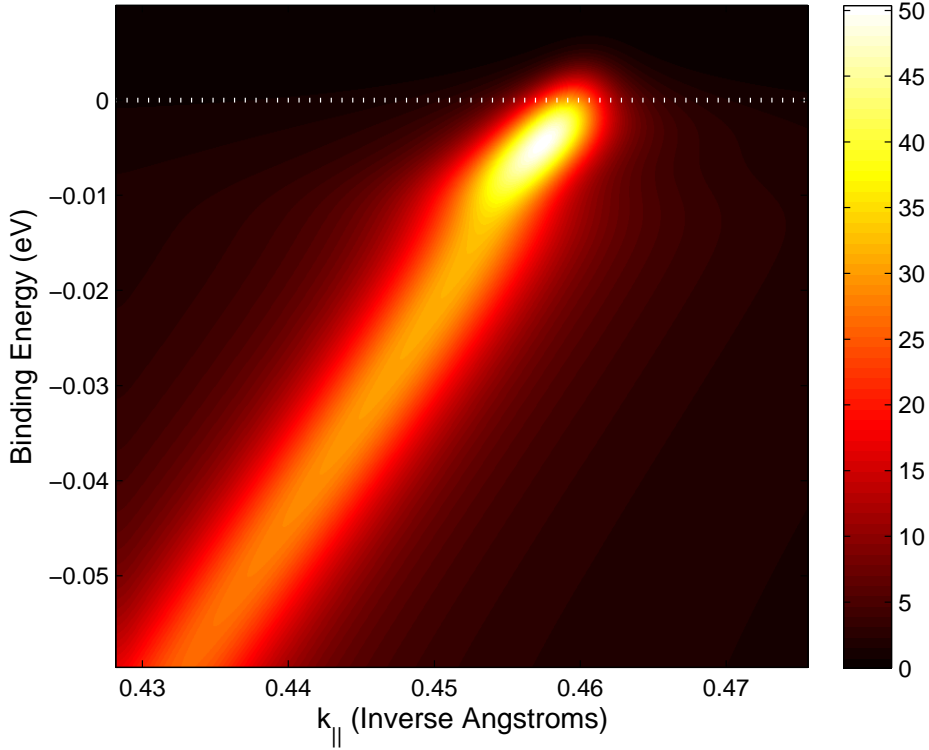}\\
  \caption{(Color online) Simulation of 8 meV kink. Relative intensities in arbitrary units are shown by the color bar on the right. With the inclusion of even modest resolution broadening mass renormalization effects are clearly much more pronounced in the imaginary versus the real part of the self energy. }\label{fig7}
\end{figure}

As an additional check to the fits of the scattering rates we have applied this kink to a simulated low temperature "nodal state". We take the lifetime to be Fermi liquid like with $\Gamma=\Gamma_{e-e}+\Gamma_{Imp}+\Gamma_{e-ph}$ and then calculate the self-consistent real part of the self energy by Kramers-Kronig transform. The bare dispersion is set to $E(k)=v_{F}(k-k_{F})$ with $v_{F}$=2.2 eVA and $k_{F}$=.46. The impurity scattering in this simulation is set to $\Gamma_{Imp}$ = 5 meV and the electron-electron lifetime is $\Gamma_{e-e}=2\beta((\pi k_{B}T)^{2}+\omega^{2})$ with $2\beta$=1.58 \cite{ref21}. $\Gamma_{e-ph}$ is calculated from Equation \ref{intgam} with $\alpha^{2}$=1.25 and a Gaussian mode of width 1 meV. The quadratic, Fermi liquid form of $\Gamma_{e-e}$ is used for simplicity and at very low energies will not deviate substantially from the low temperature cubic form deduced in \cite{ref22}. The raw spectrum is finally broadened with the energy and momentum parameters from the experiment cited above. The result is plotted in Figure \ref{fig7}. Though obviously an oversimplification of the true low energy nodal state of Bi2212, the simulation captures the essential features of the near Fermi level spectrum we are concerned with. Most importantly, we find that the renormalization of the imaginary part of the self energy in the vicinity of the kink is much more obvious and dramatic than that of the real part. While the imaginary part of the self energy appears as a drastic sharpening of the spectrum upon approaching $E_{F}$ the corresponding change in the real part of the self energy, even with a limited 5 meV energy broadening, is much weaker and will naturally appear as a much more subtle feature of the true ARPES spectra in the vicinity of the node. This is precisely what is observed in the experiment. It is thus not surprising that previous studies of Bi2212, even at reasonably high resolution, have not identified this feature in the spectrum. Certainly the reduced momentum resolution of higher energy synchrotron based ARPES studies of Bi2212, in addition to the contaminating presence of the bilayer splitting near the node, would make this feature otherwise impossible to detect.

\section{Discussion and Conclusions}

While the study of electron-phonon coupling in correlated electron materials such as the cuprates is of interest in its own right there are several ramifications as regards the more general problem of understanding the electronic structure of Bi2212. First, unlike the many high energy band renormalizations discussed in the introduction, this finding represents an unambiguous identification of such a renormalization with a particular bosonic mode. In this case the coupling of electrons in the 2D copper oxide sheets of Bi2212 to an out of plane c-axis gerade mode is clearly identified. The identification of this mode from the step in the imaginary part of the self energy is supported by the agreement between the model calculation of the scattering rate in the presence of such a mode and those measured experimentally.

While the form of $\delta_{\vec{k}}$ shown by this mode is in qualitative agreement with the prediction of Ref. \cite{ref13}, which takes a much more comprehensive approach to modeling the electron-phonon coupling in the presence of a d-wave superconducting gap, e.g. indicating that such a coupling is possible a priori, the data bares out an important additional observation. That is, a full theory of electron-phonon coupling in Bi2212, in the presence of the d-wave requires a proper treatment of the angular dependence of the kink as one scans around the Fermi surface. In the reference it was assumed that this effect would be small and one could well approximate the shift of the kink energy by pinning it the maximal d-wave superconducting gap $\Delta_{0}$. Here it is shown that the kink energy close to the node is pinned to the "local" gap value $\Delta(\phi)$.

In summary we have reported the observation of coupling to a zero momentum optical phonon.  Such coupling can be viewed as the relaxation of the lattice associated with the creation of a charged photohole in this ionic material.  We note that a similar effect has been discussed elsewhere (\cite{ref23} and references therein).  Such relaxation effects have important implications for the transport in such materials.

\begin{acknowledgments}
The authors would like to acknowledge useful discussions with Chris Homes, Phil Allen and Tony Valla. The research work described in this paper was supported by the Department of Energy under Contract No. DE-AC02-98CH10886.
\end{acknowledgments}

\bibliography{kinkbib}

\end{document}